\documentclass[9pt,twocolumn]{extarticle}
\pdfoutput=1
\usepackage{soul}
\usepackage{titlesec} 

\usepackage[superscript, nomove]{cite}

\usepackage{float}
\usepackage{caption}
\usepackage{lipsum,ulem}
\usepackage[verbose]{placeins}

\usepackage{dsfont}
\usepackage{graphicx}
\usepackage{subcaption}
\usepackage[margin=0.9in]{geometry}
\usepackage[usenames,dvipsnames]{xcolor}
\usepackage[colorlinks,linkcolor=Blue,urlcolor=Blue,citecolor=Blue]{hyperref}
\usepackage{amsmath,amssymb}
\usepackage[squaren]{SIunits}

\usepackage{mathpazo}
\usepackage{courier}
\normalfont
\usepackage[T1]{fontenc}
\usepackage{caption}
\renewcommand{\figurename}{Fig.}
\usepackage{upgreek}

\newcommand{\ad}[1]{\textsuperscript{#1}\kern-2pt}

\makeatletter
\def\blx@maxline{77}
\makeatother

\usepackage{capt-of}

\def\mytitle{An AI-driven robotic system for two-dimensional hetero-assemblies} 

\title{\vspace{0.5cm}\Huge\textbf{\textrm{\mytitle}}}

\author{\vspace{-3.0mm}\\ Xiaoxi Li,$^{1,2,3*}$ Jinkun He,$^{1,2*}$ Haojie Liu,$^{4,5*}$ Xipeng Liu,$^{4}$ Zewen Wu,$^{4}$ Jing Li,$^{4}$ Kai Zhao,$^{1,2,3}$  Shan Li,$^{1,2}$ Xingdan Sun,$^{3}$ \\Xiaoxue Fan,$^{1,2}$  Zhiren Xiong,$^{1,2}$ Xingguang Wu,$^{1,2}$ Xuanzhe Sha,$^{1,2}$ Zhili Lin,$^{1,2}$ Caixia Yang,$^{1,2}$ Luosha Han,$^{1,2}$ Jie Xu,$^{1,2}$ \\Woye Pei,$^{1,2}$  Kaining Yang,$^{3}$ Jing Zhang,$^{1,2}$ Xiaolong Feng,$^{4}$ Tongyao Zhang,$^{1,2}$ Zhu Liang,$^{4}$ Kenji Watanabe,$^{6}$ \\ Takashi Taniguchi,$^{7}$ Ming Tian,$^{8}$ Neng Wan,$^{8,9}$ Jing Zhang,$^{1,2\dagger}$ Jianming Lu,$^{3\dagger}$ Wenjing Hong,$^{4\dagger}$ Zheng Vitto Han $^{1,2,3\dagger}$}

\date{} 
\begin{document}
\twocolumn[{
\maketitle 
\vspace{-5mm}
\begin{center}
\begin{minipage}{1\textwidth}
\begin{center}
\textit{
\\\textsuperscript{1} State Key Laboratory of Quantum Optics Technologies and Devices, Institute of Optoelectronics, Shanxi University, Taiyuan 030006, China
\\\textsuperscript{2} Collaborative Innovation Center of Extreme Optics, Shanxi University, Taiyuan 030006, China
\\\textsuperscript{3} Liaoning Academy of Materials, Shenyang 110167, China
\\\textsuperscript{4} State Key Laboratory of Physical Chemistry of Solid Surfaces, iKKEM, College of Chemistry and Chemical Engineering, Xiamen University, Xiamen, China
\\\textsuperscript{5} School of Electrical and Information Engineering, Zhengzhou University, Zhengzhou, 450001, China
\\\textsuperscript{6} Research Center for Electronic and Optical Materials, National Institute for Materials Science, 1-1 Namiki, Tsukuba 305-0044, Japan
\\\textsuperscript{7} Research Center for Materials Nanoarchitectonics, National Institute for Materials Science,  1-1 Namiki, Tsukuba 305-0044, Japan
\\\textsuperscript{8} Key Laboratory of MEMS of Ministry of Education, College of Integrated Circuits, Southeast University, Nanjing, 210096, China
\\\textsuperscript{9} State Key Laboratory of Surface Physics, Key Laboratory of Micro and Nano Photonic Structures (MOE), and Department of Physics, Fudan University, Shanghai, 200433, China
\vspace{5mm}
\\{$\dagger$} Corresponding to: jzhang74@sxu.edu.cn, jmlu@lam.ln.cn, whong@xmu.edu.cn, vitto.han@gmail.com 
\\{$\star$} These authors contribute equally.
\vspace{5mm}
}
\end{center}
\end{minipage}
\end{center}

\setlength\parindent{13pt}
\begin{quotation}
\noindent 
\section*{Abstract}
{\textbf{Nanomaterials stacked on-demand, such as rotationally assembled two-dimensional (2D) van der Waals (vdW) layered compounds, provides a versatile platform for quantum simulation and the exploration of exotic electronic phases. Currently, however, such nanoassemblies remain largely confined to inefficiency, manually operated process, limiting their potential for probing emergent physical phenomena. There is a pressing need in the field for high‑precision, automated assembling techniques, especially for the scalable fabrication of 2D twistronic heterostructures. Here, we present an intelligent automation system dedicated to the fabrication of van der Waals stacks, following the state-of-the-art protocol for dry transfer of exfoliated 2D materials. The system further employs metadata generated from each automated stacking procedure to perform reinforcement learning, thereby continuously bettering its performances. As a concrete demonstration, we fabricate twisted bilayer graphene (TBLG) — known for its challenging preparation — and exhibit its unconventional superconductivity near the magic angle. Our work may pave the way for high-throughput fabrication of low-dimensional nanomaterials including twistronic heterostructures, where integrating data mining and artificial intelligence can accelerate the discovery of novel physical phenomena.}}
\end{quotation}
}]

\newpage 
\clearpage

\section*{Introduction}

Among those low-dimensional nanoassemblies, the atomically reconstructed moir\'e superlattices in van der Waals (vdW) heterostructures are known to set emergent length scales absent in the parent crystals, leading to reconstructed electronic bands and strengthened electronic correlations\cite{macdonald_moire_2011,cao_unconventional_2018,cao_correlated_2018}. These systems are nowadays widely recognized as twistronics, in which key parameters -- including the twist angle, material selection, layer number, and external tuning knobs such as displacement field, strain, and magnetic field -- can be flexibly and independently controlled\cite{andrei_graphene_2020,kennes_moire_2021}. Within a single twistronic device, one can continuously tune carrier density, dielectric screening\cite{liu_tuning_2021}, and interlayer coupling\cite{yankowitz_tuning_2019}; moreover, the nanometer-scale moir\'e periodicity is inherently compatible with techniques such as scanning tunneling microscopy (STM)\cite{li_observation_2010,kerelsky_maximized_2019}, transport\cite{stepanov_untying_2020}, and optical spectroscopies\cite{tang_simulation_2020,xu_correlated_2020,shimazaki_strongly_2020,chu_nanoscale_2020}. These distinctive attributes have rapidly established twistronics as a leading platform for quantum simulation. Indeed, recent advances in engineering of symmetry-breaking and Berry-phase–driven nontrivial topological effects in vdW twistronic devices have unveiled a rich spectrum of quantum phases, including superconductivity\cite{cao_unconventional_2018,lu_superconductors_2019}, Mott-Hubbard insulators\cite{cao_correlated_2018,chen_evidence_2019}, moir\'e excitons\cite{tang_simulation_2020,xu_correlated_2020,shimazaki_strongly_2020,huang_correlated_2021}, Chern insulators\cite{sharpe_emergent_2019, serlin_intrinsic_2020,park_observation_2023,lu_fractional_2024}, and various other exotic correlated states\cite{regan_mott_2020, xu_correlated_2020}.

Despite rapid advances, the fabrication of vdW twistronic devices still largely relies on manual twist alignment of two or more layers -- a labor-intensive, low-throughput process that is inherently difficult to scale and prone to variability. State-of-the-art workflows typically involve pre-cut target flakes (e.g., by laser-beams \cite{park_flavour_2021}, or scanning-probe lithography \cite{li_electrode-free_2018}), followed by sequentially picking up each section of the as-cut layer with polymer stamps such as polydimethylsiloxane (PDMS) or polycarbonate (PC) \cite{cao_correlated_2018}. Stacking then proceeds via wetting-detaching at the contact wavefront between the stamp and the substrate, which requires delicate manual control and is vulnerable to interfacial contamination, superlubricity-induced slippage, and defect formation\cite{kazmierczak2021strain, uri2020mapping, lau2022reproducibility, song2018robust}. Consequently, imperfections such as bubbles, wrinkles and ruptures frequently take place, leading to notably low sample yields -- reported, for example, to be below 1/20 in certain flat-band device studies. 

Given these fabrication limitations, vast parameter spaces of moir\'e superlattices might remain uncharted, hindering the potential to uncover emergent physical phenomena. A promising route to overcome these obstacles is to scale up the fabrication of twistronic vdW heterostructures in a high-throughput, artificial intelligence (AI)-assisted manner \cite{zhao2025automated, masubuchi2018autonomous, mannix2022robotic, wang2023scientific, chen2023fully, masubuchi2020deep, masubuchi2019classifying, bencherif2024automated, lu2024machine, han2020deep, saito2019deep, greplova2020fully, sabattini2025towards, uslu2024open, ramezani2023automatic, lu2020coupling}. In recent years, early attempts have indeed been made toward robotic 2D-flake-identification and -stacking, including automaton systems operating in inert atmospheres \cite{masubuchi2018autonomous} and "four-dimensional" vdW stacking platforms for large-area chemically vapor-deposited materials \cite{mannix2022robotic}. Nevertheless, existing approaches are either inadequate for constructing complex, twist angle-critical heterostructures or lack the capacity for sustained self-optimization. To date, a dedicated platform that integrates high-precision automated stacking with evolutionary data-driven learning -- specifically designed for advanced twistronic heterostructures -- has yet to be realized. 

In this work, we report an AI-driven stacking robot capable of synergically executing state-of-the-art twistronic fabrication protocols. Utilizing computer vision to identify vdW flake, the system performs vdW layer recognition, optical focus tracking, and real-time monitoring of Newton’s ring wavefront motion, thereby accomplishing user-defined assembly tasks for custom twistronic heterostructures. Over approximately 200 hours of operation, the robot fabricated 100 hexagonal boron nitride (h‑BN) encapsulated twisted bilayer graphene (TBLG) heterostructures. Among these, about $\sim$52.4\% of the stacks achieved twist‑angle accuracies better than 0.1$^{\circ}$, and roughly $\sim$49.5\% exhibited lateral displacement errors below 2 $\mu$m. Importantly, each automated stacking cycle generates comprehensive metadata that supports a reinforcement learning loop, enabling continuous improvement in motion control, alignment precision, yield, and throughput. This work represents not only an automated nanoassembly technique but also an early prototype of a fully autonomous robotic manufacturing platform for low-dimensional materials, opening new opportunities for scalable production of twistronic devices, artificial heterostructures, and designer quantum material systems.

\begin{figure*}[t!]
	\centering
	\includegraphics[width=0.88\linewidth]{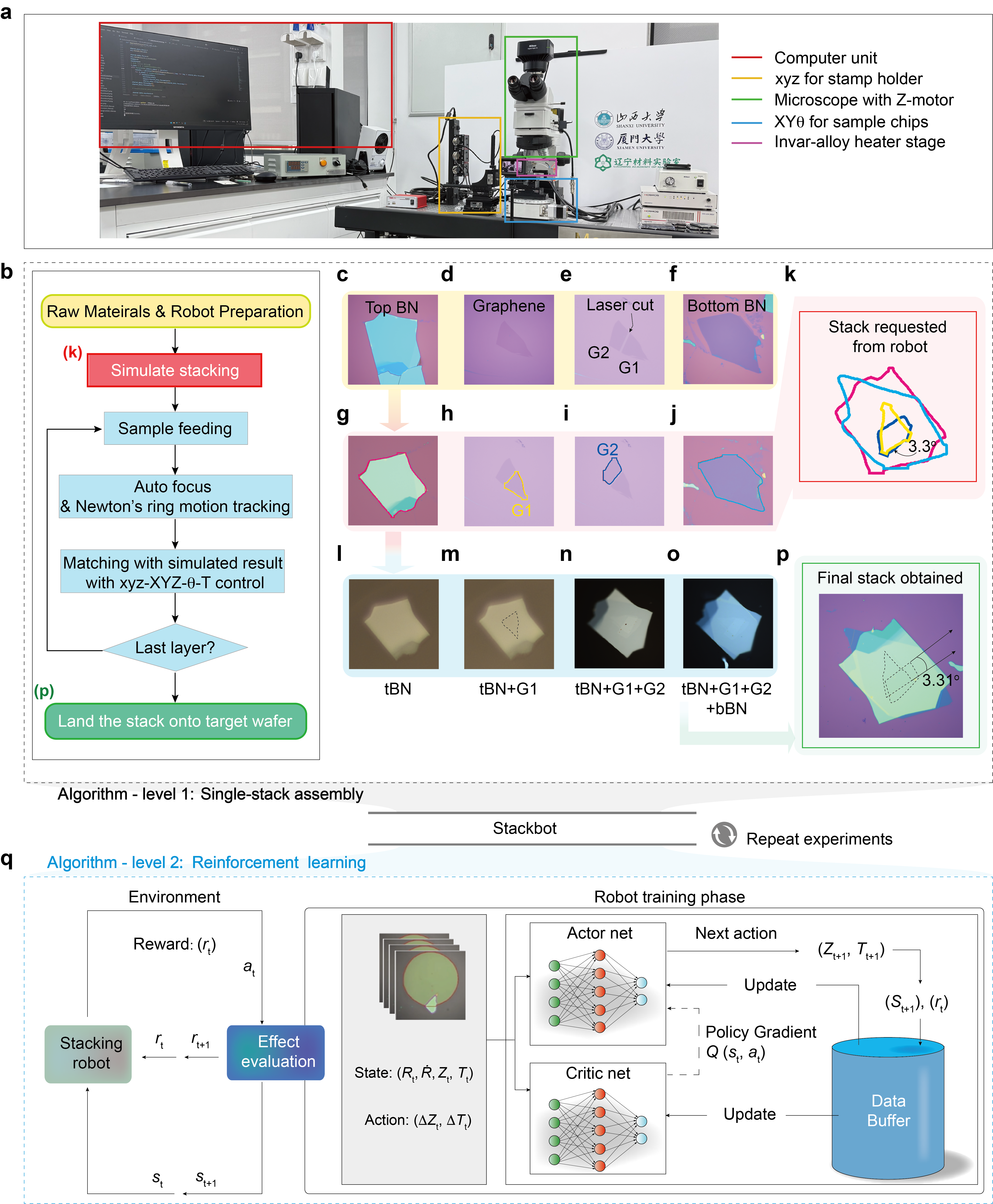}
	\caption{\textbf{AI-assisted automation for the assembly of twistronic 2D materials.} (a) Photograph showing the outfit of the general assembly of the machine. (b) Level-1 workflow for a single stack: the controller first simulates the target heterostructure from AI-recognized flake outlines (c-k). The system then performs auto-focus and Newton's-ring tracking during approach (see Supplementary Video 1), while the PDMS stamp (\(xyz\)), the sample wafer stage (\(XY\theta\)), the microscope focus (\(Z\)), and temperature (\(T\)) registers the live view to the simulated plan. The loop repeats layer-by-layer until completion (l-o), after which the stack is landed on the target wafer (p). (q) illustrates the so-called Level-2 algorithm for reinforcement learning. }
\end{figure*}

\begin{figure*}[t!]
	\centering
	\includegraphics[width=0.9\linewidth]{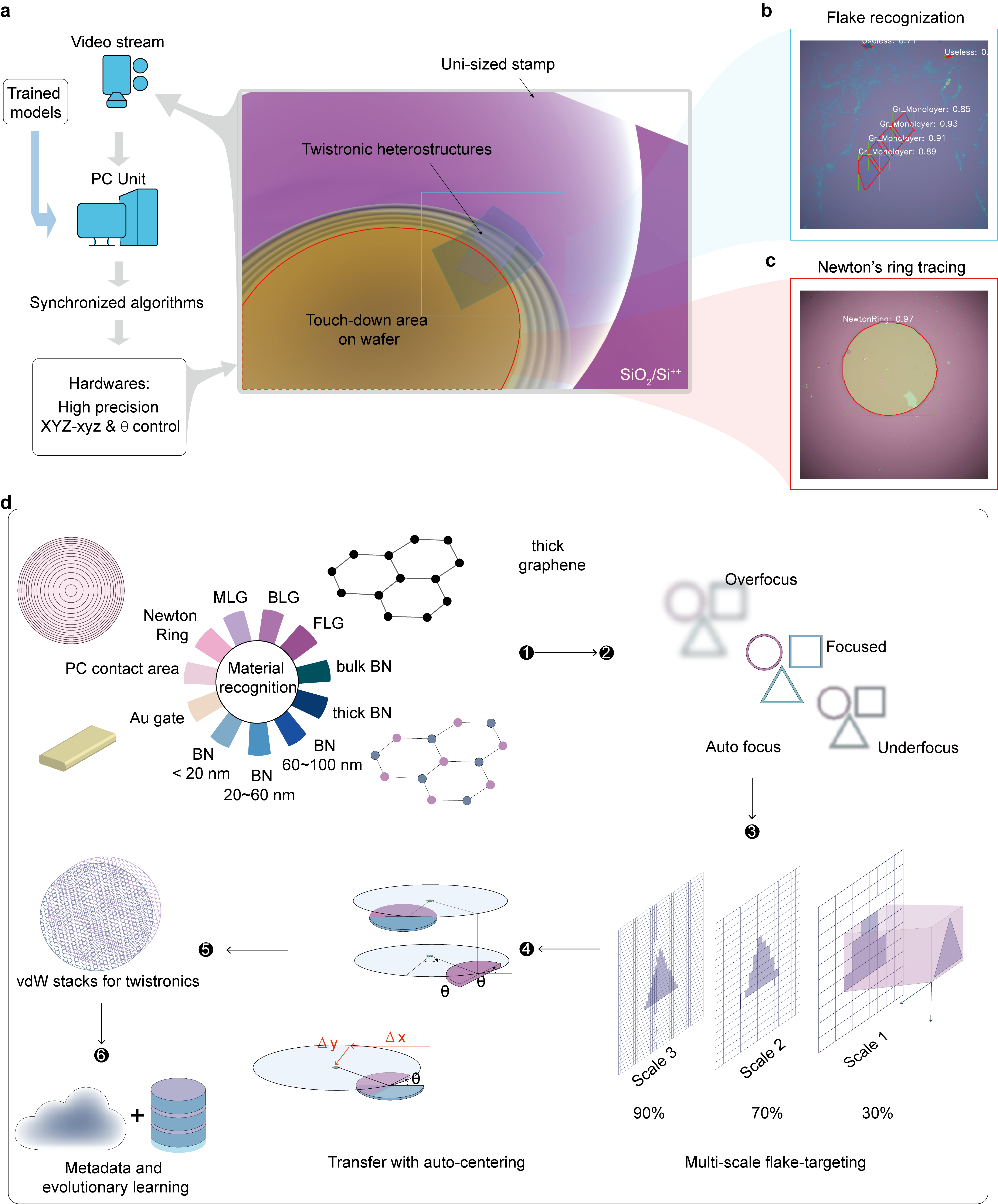}
	\caption{\textbf{Pipelines of the synchronized algorithm and hardware in the stack-bot.} (a) General view of the strategy of the stack-bot: machine-vision models are trained and fed to the program, while the algorithms drive hardware to execute multi-layered twistornic stacks as aimed The real-time monitoring of video stream is then sent back into the algorithm loop. Illustrations of (b) vdW flake recognition and (c) Newton's ring tracing during the loop in (a). (d) Schematic of a pipeline of the detailed strategies applied in (a). First, material recognition of different vdW layered compound on SiO$_{2}$ wafers are realized through the model training, and the auto-focus section tackles the visual tasks during the whole stacking process. A multi-scale flake-targeting method, as well as auto-centering of chips, is further used. The parameter sapce of all hardware and videos of the prepared vdW stacks are then registered in the metadata for reinforcement learning.}
\end{figure*}

\begin{figure*}[ht!]
	\centering
	\includegraphics[width=0.9\linewidth]{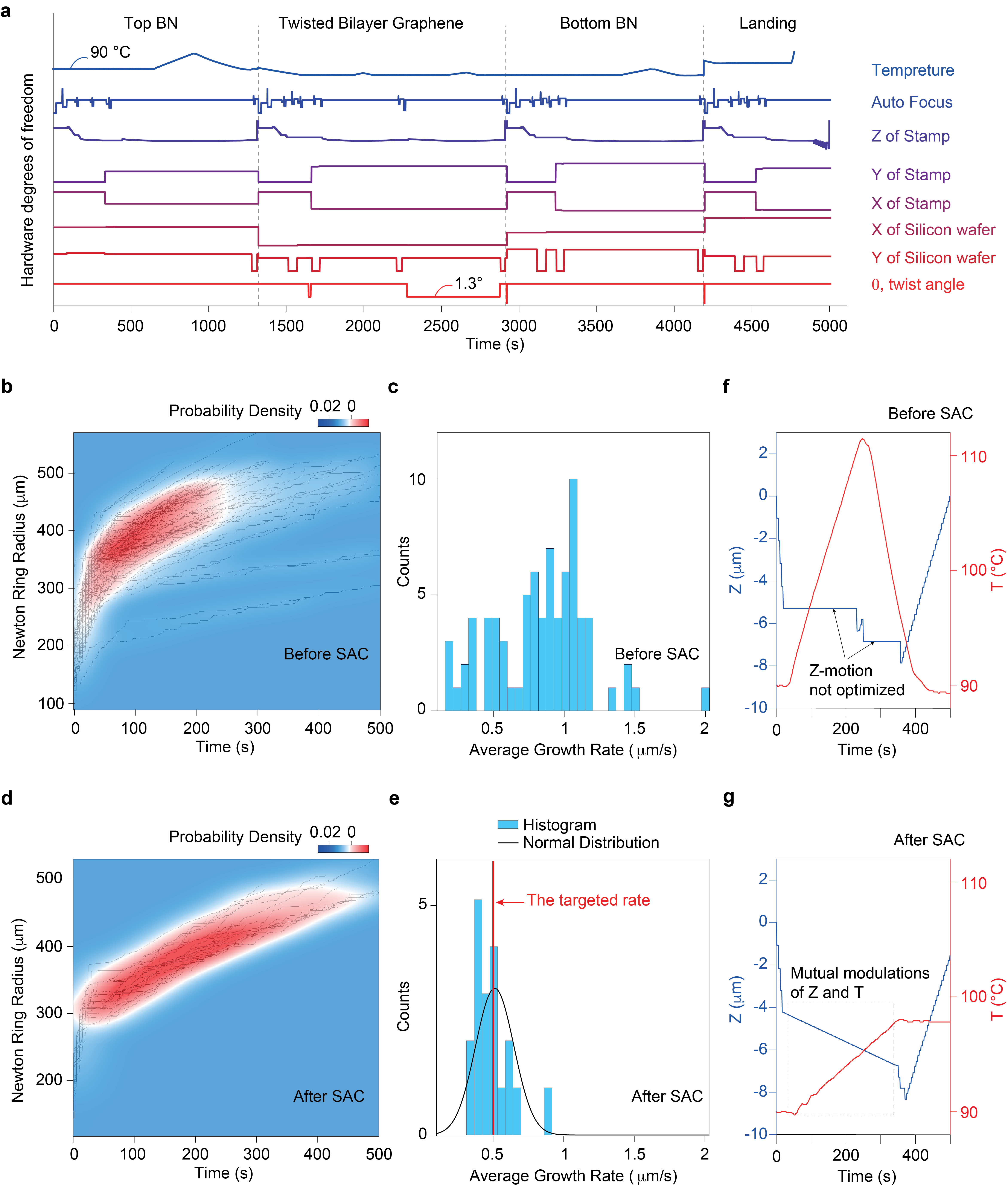}
	\caption{\textbf{Reinforcement learning based  on the metadata of vdW stacks.} (a) Data logged in a single loop of stack assembly. All the XY-xyz-$\theta$-Z$_{\textrm{optical}}$-$T$ degrees of freedom are recorded along with the time axis, also accompanied with a video record in the metadata. (b) Probability density plotted for the Newton's ring radius as a function of time, for a statistics of $\sim$ 70 stacks. After application of the secondary reinforcement learning (the SAC algorithm), the probability density of the Newton's ring radius as a function of time are narrowly distributed in (c), with better control of the growth rate (i.e., the PDMS wavefront motion speed). Histogram of sample counts in their growth rates are plotted in (d) before SAC, and (e) after SAC, respectively. (f) and (g) illustrate typical curves of Z-motion and T-control during the auto-stacking process before and after SAC, respectively.
	}	
	\label{fig:fig3}
\end{figure*}

\section*{Results}
\noindent\textbf{Experimental setup of the AI-driven robot for vdW twistronics.} We begin by describing the system architecture, operational workflow, and overall performance. A photograph of the benchtop implementation is provided in Figure 1a. To incorporate the latest manual transfer protocols into our robotic platform, the overall workflow was divided into two main stages. First, raw materials are prepared via exfoliation. Computer vision is then applied to extract the geometric fingerprints—including shape, size, and distinctive features—of each flake. Subsequently, the second stage, namely the automated assembly of a single heterostructure, is executed according to the flowchart shown in Figure 1b. Further hardware details are provided in the Methods section. The machine identifies van der Waals flakes through a hierarchical computer‑vision pipeline that combines a YOLO‑based object‑detection algorithm for coarse localization (Supplementary Figure 1) and a Continuous Refinement Model (CRM) for fine‑grained feature extraction\cite{shen2022high} (Supplementary Figure 2). These perception modules are coordinated by a custom Python software package (Supplementary Figure 3), which constitutes the primary algorithmic layer (Level 1).

We evaluate our Level‑1 algorithm using the widely studied example of twisted bilayer graphene (TBLG). Figures 1c–f present representative raw materials provided to the robot: a top h‑BN (tBN) capping layer, monolayer graphene that has been laser‑partitioned into sub‑flakes G1 and G2 (sharing the same crystallographic orientation), and a bottom h‑BN (bBN) substrate. The perception pipeline detects each flake and returns its outline (Figures 1g–j). From these contours, our in-house software generates a stacking‑placement plan (Figure 1k),  specifying desired layer overlap and a target twist angle (e.g. \(3.3^{\circ}\)), and then produces corresponding displacement- and angle‑control commands. Figures 1l–o showcase frames captured during a fully-automated assembly: tBN \(\rightarrow\) tBN+G1 \(\rightarrow\) tBN+G1+G2 \(\rightarrow\) tBN+G1+G2+bBN. The resulting heterostructure (Figure 1p) exhibits a measured twist angle of $\sim$ \(3.3^{\circ}\) (determined from optical micrographs using image‑analysis software; see Methods), consistent with the planned value in Figure 1k. Batch statistics over 100 stacks are summarized in Extended Data Figure 1, quantifying the accuracy and repeatability: the distribution of absolute twist-angle errors is dominated by sub-\(0.3^{\circ}\) events with a small tail exceeding \(0.5^{\circ}\). We note that lateral registration errors on the order of micrometers (\(\mu\)m) occasionally occur, which may originate from slippage induced by electrostatic forces and/or superlubricity effect \cite{uri2020mapping,song2018robust}. Collectively, these results demonstrate that real‑time vision, wetting‑front tracking, and integrated control of XY-xyz-$\theta$-Z$_{\textrm{optical}}$-$T$ enable reliable, high-fidelity assembly of twist-critical vdW stacks.

As shown in Figure 1r, the system implements a secondary algorithmic layer (Level 2) for continuous self-improvement. Upon completion of each automated stacking cycle, a suite of metadata -- such as image-registration residuals, contact-quality scores, drift estimates, and other operational parameters including temperature and stack-motion profiles (see Supplementary Repository Data 1) -- is evaluated and fed to a deep reinforcement learning framework based on the Soft Actor–Critic (SAC) architecture. Within this framework,  the actor network proposes refined actions (e.g., updated vision thresholds, dwell times, or micron‑scale translations), while the critic updates a value function using historical operation data (data buffer). This closed-loop, data-driven earning process is one of the central results of this study: it optimizes stacking policies through iterative learning across multiple experiments, mitigating limitations in vdW material stacking arising from geometric variations, air bubbles, or contamination-defects, which will be discussed in the following sections.

\bigskip
\noindent\textbf{Synergization of AI-algorithms.} To elaborate in more details on the software and working principle of the vdW stacking robot, we present a general schematic of the stack-bot in Figure 2a. Once trained models are loaded into the program, the closed loop of "PC program/algorithms --  hardware with 8 degrees of freedom (XY-xyz-$\theta$-Z$_{\textrm{optical}}$-$T$) -- video stream" is executed to assemble the designed multilayer twistronic stacks (as specified in Figure 1k). A full list of hardware components can be found in Supplementary Table 1. To ensure consistency across batches, PDMS stamps are prepared using a hot‑casting method that yields uniform stamp sizes\cite{lin2026batchfabricated}. The key handling part is the touching point of the PDMS wavefront on the SiO$_{2}$ wafer, highlighted by the blue box in Figure 2a. The algorithm adjusts the wetting line (red solid line in Figure 2a) to fully cover the target vdW flake, then retracts the stamp to pick it up,  while dynamically coordinating the $Z$ $\&$ $T$ parameters. Representative examples of vdW flake recognition and Newton’s‑ring pattern detection are shown in Figures 2b and 2c, respectively.

Figure 2d outlines the algorithm pipeline of the stack-bot. First, machine-vision models are trained on different materials, including few layered (FL) h-BN, FL-graphene, electrodes, and the Newton's ring patterns at the wetting-detaching wave front of PDMS stamps  (more data can be seen in Supplementary Table 2 and Supplementary Figures 4-6). An auto‑focus module then supports continuous machine‑vision tasks throughout the stacking process (marked as Step 2; see also Supplementary Note 1). Subsequently, a multi‑scale flake‑targeting method (Step 3; Supplementary Figure 7) and an automated chip‑centering routine (Step 4; Supplementary Note 2) are employed to achieve the designed vdW twistronic stack (Step 5). All relevant hardware parameters and video recordings from the stacking procedure are registered as metadata to support reinforcement learning (Step 6). An alternative visualization of this pipeline is presented in Extended Data Figure 1, which includes detailed schematic diagrams of the algorithm.

\bigskip
\noindent\textbf{Reinforcement learning of the autonomous twistronic assembling process.} We now analyze the Level‑2 reinforcement learning implemented in our autonomous twistronic assembly system. Figure 3a depicts the metadata logged during a single assembly loop of a four‑layer van der Waals stack—consisting of top h‑BN, twisted bilayer graphene, and bottom h‑BN—transferred onto a SiO$_{2}$ substrate. All mechanical degrees of freedom (XY-xyz-$\theta$-Z$_{\textrm{optical}}$-$T$) are recorded along with the time axis. It is seen that the variable changes in each hardware are recorded along the time axis, capturing the variations in each hardware parameter throughout the process, which completes within 90 minutes. Some key parameters such as sample stage temperature and twisting angle are highlighted in the corresponding traces. The video stream from each stacking cycle is also stored as part of the metadata (see Supplementary Video 1). To illustrate the impact of reinforcement learning, we first plot the probability density for the Newton's ring radius as a function of time across $\sim$ 70 stacking trials (Figure 3b). It is seen that the traces of wavefront motion of the Newton's ring are scattered with a broad probability density in Figure 3b, which is also reflected by the wide distribution in its histogram in Figure 3c. 

A key advantage of our stack-bot system, as noted earlier, lies in its Level‑2 reinforcement learning capability, which leverages metadata collected across multiple stacking cycles. Through this accumulating dataset, the system progressively refines its own operational performance. To illustrate this learning capacity, we examine the optimization of the moving velocity (or growth rate) of the Newton’s‑ring wavefront as a representative case. A Soft Actor-Critic (SAC) agent \cite{haarnoja2018soft} is implemented (see Supplementary Note 3) based on the statistics of approximately 70 stacks in Figure 3a, and deployed to navigate the high-dimensional parameter space of the system (Supplementary Figure 8). By continuously assessing irregularities in the propagation of the Newton’s‑ring wavefront, the SAC agent dynamically coordinates key hardware degrees of freedom—most notably optimizing the real‑time coupling between the stamp’s vertical position ($Z$) and the substrate temperature ($T$). This closed‑loop adaptive control progressively tunes the stacking parameters, stabilizing the average growth rate around the desired target value.

Following the implementation of secondary reinforcement learning via the SAC algorithm, the temporal evolution of the Newton’s‑ring radius exhibits a notably narrower probability distribution (Figure 3d), reflecting improved control over the growth rate—that is, the speed of the PDMS stamp’s wavefront motion. Consistent with this, the histogram of growth‑rate samples in Figure 3e shows a pronounced concentration around the target value of 0.5 $\mu$m/s (marked by the solid red arrow), as predefined in the SAC parameters. Further insight is provided by zoomed‑in profiles of the PDMS stamp’s $Z$‑motion and the stage temperature $T$ recorded during automatic stacking, before and after SAC deployment (Figures 3f and 3g, respectively). In the highlighted region of Figure 3g, the simultaneous, coordinated adjustment of $Z$ and $T$ is evident, enabling a stable and well‑regulated growth rate. These results confirm that the stack‑bot’s reinforcement‑learning framework effectively enhances performance through iterative data accumulation. 

Looking forward, the system is inherently capable of acquiring more advanced operational skills—for instance, identifying optimal paths to circumvent local contaminants (e.g., particles or bubbles), determining sample‑specific ideal landing angles or speeds, and self‑optimizing toward faster overall stacking processes. Datasets documenting system performance before and after the secondary reinforcement‑learning stage are provided in Supplementary Figures 9–12 for detailed comparison.

\begin{figure*}[ht!]
	\centering
	\includegraphics[width=0.88\linewidth]{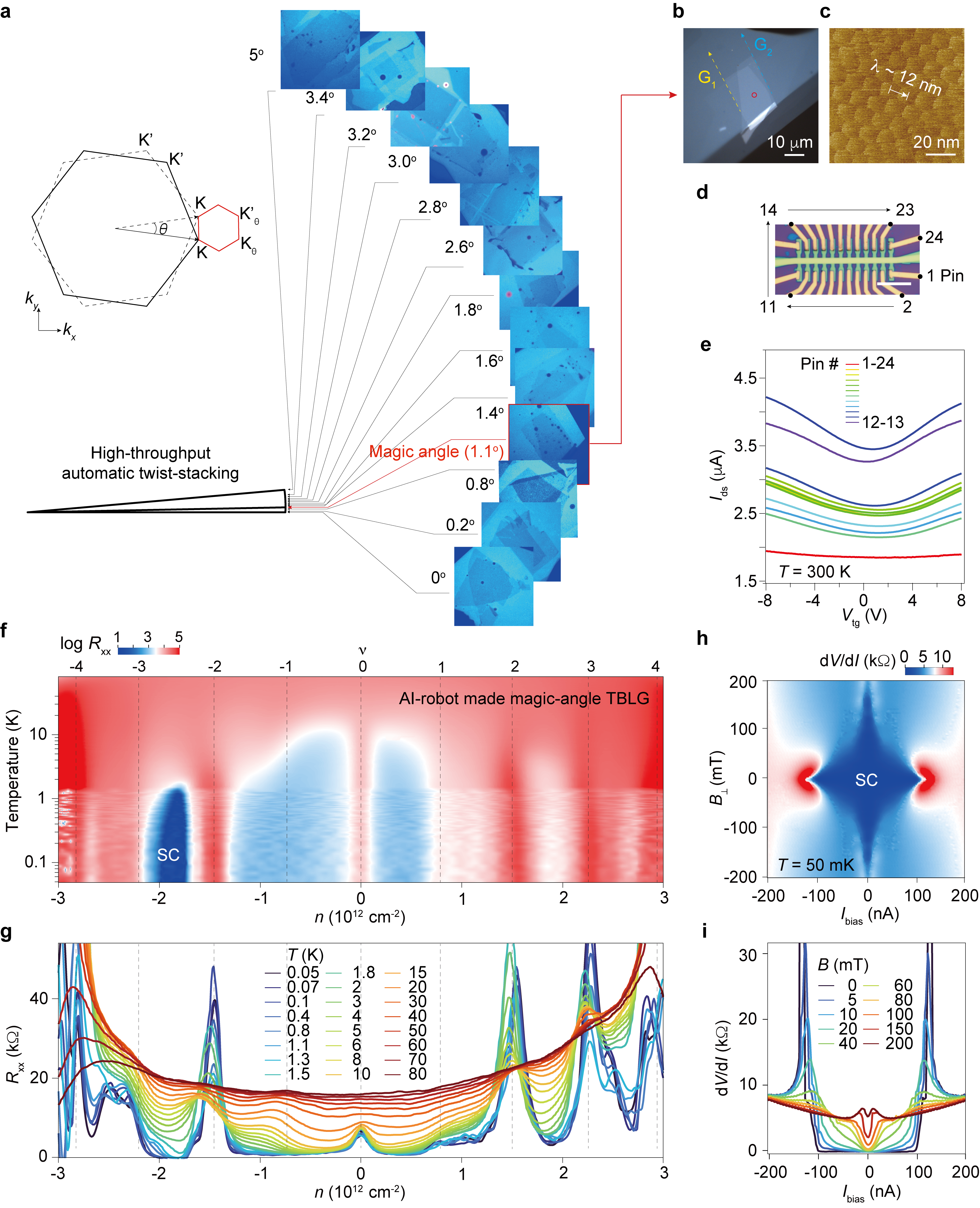}
	\caption{\textbf{High-throughput fabrication of twistronic vdW heterostructures.}
		(a) Demonstration of high-throughput auto-stacking of h-BN encapsulated TBLG over a span of twist angles from $0^\circ$ to $5^\circ$. The inset shows the mini-Brillouin zone formed by the moir\'e superlattice.
		(b) Micrograph of a typical stack with a twist angle of $1.1^\circ$, before encapsulation by the bottom h-BN.
		(c) The stack in (b) inspected by atomic force microscopy (AFM). The moir\'e pattern in real space has a wavelength of $12\,\mathrm{nm}$, consistent with the expected value for $1.1^\circ$.
		(d) Optical image of a typical device fabricated from the auto-assembled stack, with pin numbers marked.
		(e) Two-terminal field-effect curves measured across different pins at $300\,\mathrm{K}$.
		(f,g) Color maps of field-effect curves as a function of temperature. Line cuts are shown in (g). A clear superconducting region is observed on the hole side, near filling $\nu=-3$.
		(h,i) Color maps of $\mathrm{d}V/\mathrm{d}I$--$I_{\mathrm{bias}}$ curves as a function of magnetic field. Line cuts at different fields highlight the superconducting behavior.}
	\label{fig:fig4}
\end{figure*}

\bigskip
\noindent\textbf{Scalable fabrication of twistronic quantum simulators.} The center of this study is the application of such an autonomous stacker in the field of condensed matter physics. We now study the case of high-throughput fabrications of twisted bilayer graphene (TBLG). Figure 4a displays optical micrographs of typical hexagonal boron nitride (h-BN)-encapsulated TBLG stacks, fabricated across a twist-angle range of 0 to 5$^{\circ}$ (additional stacks are provided in Supplementary Figures 13–14; all optical images are false-colored for clarity). 

The inset in Figure 4a illustrates the mini-Brillouin zone (red) resulting from the rotation (from solid to dashed black lines) of the original graphene Brillouin zones. Prominent flat-band phenomena are expected near the so-called "magic angle" of approximately 1.1$^{\circ}$, which requires a stringent fabrication tolerance of about $\pm$ 0.2$^{\circ}$. We therefore focus our detailed characterization on this narrow angular window, highlighted by the red box in Figure 4a.

To directly probe the moiré superlattice formed in real space, atomic force microscopy (AFM) was performed on a representative stack prior to final h-BN encapsulation (Figure 4b). Figure 4c shows an AFM image within the region marked by the red circle in Figure 4b, revealing a moir\'e wavelength ($\lambda$) of $\sim$ 12 nm over a 1 $\times$ 1 $\mu$m$^{2}$ area (see Supplementary Figure 15). This measured period agrees well with the expected $\lambda$ $\sim$ 12.8 nm \cite{macdonald_moire_2011} for a twist angle of 1.1$^{\circ}$. Among six TBLG stacks targeted at 1.1$^{\circ}$ (see Supplementary Figure 15-16), the typical angular error $\delta \theta$ between the target angle and the value extracted from AFM mior$\acute{\textrm{e}}$ patterns was $\sim \pm$ 0.5$^{\circ}$, as summarized in Figure 4d. Notably, TBLG samples exhibiting significant moiré inhomogeneity (Supplementary Figure 17) or exceptionally weak interlayer coupling—which precluded the detection of moiré fringes—were excluded from our statistical analysis.

The fabricated TBLG stacks were then processed into Hall-bar geometry devices, each featuring a gold top gate (see Methods for fabrication details). We note that among the stacks near the magic angle, a slight interlayer rotation could occur during the final h-BN encapsulation and transfer steps. For instance, in Device S-1.3-1, while AFM measurement prior to bottom h-BN encapsulation indicated a moir\'e pattern corresponding to a twist angle of approximately 1.3$^{\circ}$, electrical characterization later revealed a twist angle closer to 1.1$^{\circ}$. This device was subsequently characterized systematically at low temperatures. An optical micrograph of the device is shown in Figure 4e, with contact pins labeled 1–24. Room-temperature two-terminal field-effect measurements across various contact pairs demonstrated homogeneous transport behavior (Figure 4f), consistent with previous reports\cite{cao_correlated_2018, PhysRevLett.129.186801}. The device was then cooled down in two separate cycles—with base temperatures of 1.5 K and 50 mK, respectively—for detailed transport measurements (further data provided in Supplementary Figures 18–23).

Figure 4g reveals the emergence of correlated insulating states (measured between contact pins 11 and 12 in Device S-1.3-1) at integer electron fillings per moir\'e unit cell below a few tens of kelvin. Upon further cooling, certain doping regions exhibit even lower resistance, as indicated by the blue areas in the map. Near 1 K, a superconducting pocket begins to develop on the high-density side of the $\nu$ = -3 filling, labeled “SC” in Figure 4g. Corresponding line profiles of the longitudinal resistance $R_{\textrm{xx}}$ versus carrier density $n$ at selected temperatures are shown in Figure 4h, where $n$ is calibrated via Hall coefficients (Supplementary Figure 18); the observed integer fillings align well with a system of the twist angle of 1.1$^{\circ}$ (see Methods). In Figure 4i, the differential resistance d$V$/d$I$ around $n \sim$ -1.8 $\times$ 10$^{12}$ cm$^{-2}$ is plotted as a function of bias current $I_{\textrm{bias}}$ and out-of-plane magnetic field $B_{\perp}$. Further line cuts of d$V$/d$I$ versus $I_{\textrm{bias}}$ at fixed $B_{\perp}$ (Figure 4j) demonstrate characteristic superconducting behavior.

In addition to pins 11–12, measurements across different contact pairs at 50 mK revealed slight variations in the superconducting features. For example, pins 8–9 exhibited two additional superconducting pockets on the electron-doped side, as shown in Extended Data Figure 3. The differential resistance d$V$/d$I$ within these three pockets (SC1, SC2, and SC3) was further mapped in the parameter space of $I_{\textrm{bias}}$ and $B_{\perp}$ (Extended Data Figure 4). Extended Data Figure 5 reflects subtle spatial inhomogeneity, suggesting that the observed flat-band superconductivity is highly sensitive to local variations. 

The successful fabrication and electronic characterization of this magic-angle TBLG device demonstrate the high reliability and precision of our stacking robot in producing sophisticated van der Waals heterostructures for advanced twistronic research. Our approach further points toward a future in which nanoassembly of van der Waals materials and other low-dimensional systems may transition from artisanal practice to programmable manufacturing. By coupling machine vision with autonomous robotic control, it outlines an early framework for closed-loop, data-driven construction of designer heterostructures. Such capability may ultimately enable scalable realization of complex twistronic and quantum material platforms that are currently beyond the reach of manual fabrication.

\bigskip

In summary, we have developed an AI‑driven robotic platform dedicated to the fabrication of twistronic quantum devices. Our system executes vdW layer recognition, optical focal‑plane tracking, and real‑time monitoring of Newton’s‑ring wavefront motion throughout the pick‑and‑place workflow. All XY-xyz-$\theta$-Z$_{\textrm{optical}}$-$T$ degrees of freedom, as well as a video stream of each stacking cycle, are continuously logged. This comprehensive metadata collection enables a reinforcement‑learning framework that grants the system autonomous self‑optimization capability. Specifically, we show that a Soft Actor‑Critic agent can effectively learn to regulate the complex dynamics of Newton’s‑ring propagation during automated vdW flake transfer. We further demonstrate the high‑throughput fabrication of TBLG stacks. In particular, magic‑angle TBLG devices with a twist angle of $\sim$ 1.1$^{\circ}$ exhibit hallmark flat‑band phenomena, including superconductivity and correlated insulating states, confirming the high reliability and precision of our stacking robot in assembling complex twisted vdW heterostructures. Our work opens new avenues for large‑scale exploration in twistronics, promoting data‑driven and AI‑assisted workflows that alleviate the reliance on manual assembly and may accelerate the discovery of emergent quantum phases.

 \clearpage

\section*{Methods}
\vspace{3mm}
\noindent\textbf{Hardware of the robot.} Muti-degrees of freedoms are adopted, including xyz-axis for the PDMS stamp (to pick up the vdW flakes), XY-axis for wafer chip, and a rotation stage to control the twist angle $\theta$ with precision of better than 0.01 degree, also the microscope is equipped with a Z axis, as well. Heater stage is equipped with Invar alloy, to avoid possible thermal drifts/vibrations during experiments.

\vspace{3mm}
\noindent\textbf{Materials preparations.} High-quality graphene and hexagonal boron nitride flakes were mechanically exfoliated from bulk crystals onto SiO$_{2}$/Si substrates. To facilitate the yield of large-area monolayer graphene, the substrates were pretreated with O$_{2}$ plasma (300 W, 15 s) prior to exfoliation. For precise twist-angle assembly, graphene flakes were patterned via a laser-assisted isolation process employing a 532 nm picosecond pulsed laser system, resulting in graphene segments separated by a typical spacing of approximately 5 $\mu\text{m}$. The well-defined straight edges produced during laser patterning were preserved as geometric references for the subsequent optical determination of the relative twist angle.

The assembly of twisted bilayer graphene heterostructures was conducted under ambient conditions via an automated robotic dry-transfer process. This process utilized a poly (bisphenol A carbonate) (PC)/polydimethylsiloxane (PDMS) stamp supported on a glass slide. The PDMS stamps were fabricated using a batch hot-casting droplet method \cite{lin2026batchfabricated}to ensure controlled geometry and superior surface quality.

\vspace{3mm}
\noindent\textbf{Atomic force microscopy measurements.} For the twisted bilayer graphene near the magic angle ($\sim$ 1.1$^{\circ}$ ), the moiré superlattice was characterized prior to the final top hBN encapsulation using a Bruker Dimension Icon atomic force microscope (AFM). The measurements were performed in lateral force microscopy (LFM) mode to resolve the moiré fringes, utilizing FMV-A probes (Bruker, nominal spring constant $\sim$ 2.8 N/m). Beyond verifying the moiré periodicity and twist-angle uniformity, this mapping process served as a spatial guide, allowing us to pinpoint optimal regions for subsequent electrode patterning and device fabrication.

\vspace{3mm}
\noindent\textbf{Nano fabrications.} Following the automated stacking, standard lithography and metallization processes were performed to define both the contact electrodes and the top-gate structure. Electron beam lithography (EBL) was carried out using a Zeiss Sigma 360 scanning electron microscope (SEM) equipped with a Raith Elphy Quantum pattern generator. To establish electrical connections, one-dimensional (1D) edge contacts were fabricated via reactive ion etching (SAMCO, RIE-10NR) followed by electron beam evaporation (Angstrom, Amod). The thicknesses of the Cr/Au metal stacks were 5/30 nm for the top-gate electrodes and 5/50 nm for the contact electrodes, respectively.

\vspace{3mm}
\noindent\textbf{Electrical measurements.} Initial device screening was conducted at room temperature in a two-terminal configuration using a KEYSIGHT B1500 Semiconductor Device Analyzer. For detailed magneto-transport studies, the devices were loaded into an Oxford cryostat with a base temperature of 1.5 K, equipped with a perpendicular magnetic field of up to 12 T. Four-terminal longitudinal (R$_{xx}$) and Hall (R$_{xy}$) resistances were measured using a standard low-frequency lock-in technique. An AC excitation current of $I_\mathrm{bias}$=100 nA was applied at a frequency of 13.333 Hz using a Stanford Research Systems SR860 lock-in amplifier. Subsequently, measurements were extended to a base temperature of 50 mK in a Bluefors dilution refrigerator. In this ultra-low temperature regime, the AC excitation current was reduced to $I_\mathrm{bias}$=1 nA at 13.333 Hz. These high-sensitivity measurements were carried out using a Stanford Research Systems SR830 lock-in amplifier. Throughout the cryogenic measurements, A DC voltage was applied using a Keithley 2400 Source Measure Unit (SMU) under a $\sim1$ nA current bias.

\vspace{3mm}
\noindent\textbf{Twist Angle Determination.} Hall measurements were performed at $1.5~\mathrm{K}$ under a perpendicular magnetic field of $0.5~\mathrm{T}$. In the linear Hall regime with a single dominant carrier type (Supplementary Figure18), the Hall resistance follows $R_{xy}=B/(n_{H}e)$, enabling a direct calibration of carrier density. Due to residual doping, the charge neutrality point (CNP) is shifted from zero gate voltage and located at $\sim-3.75$ V; all carrier densities were therefore referenced symmetrically to the CNP. Using this CNP-symmetric Hall calibration, the carrier density corresponding to a fully filled moir'e superlattice unit cell (filling factor $\nu=4$) is determined to be $\sim2.649\times10^{12}$ cm${^{-2}}$. The unit-cell area is given by $(\sqrt{3}/2)(a/\theta)^2$, where $a$=0.246 nm is the graphene lattice constant. Combining these relations yields the twist angle $\theta=\sqrt{n(\sqrt{3}/8)a^2}$, from which the twist angles extracted for different pins of the 1.3-1 device fall in the range $1.067^\circ$–$1.187^\circ$.

\vspace{3mm}
\noindent\textbf{AI algorithms.} The AI algorithms consist of three core functional modules, synergistically enabling automated and high-precision twistronic fabrication. First, a computer vision-based vdW flake recognition module, which achieves real-time segmentation and localization of vdW flakes (e.g., h-BN and graphene). Second, an adaptive optical focus tracking algorithm combines phase-correlation analysis and deep learning-based feature matching to dynamically adjust the microscope’s Z-axis position, compensating for tiny vibrations or sample height variations and maintaining stable focus on the flake surface. Third, a reinforcement learning (RL)-driven motion control module processes metadata from each stacking cycle to update action policies for the xyz-axis PDMS stamp and temperature. This evolutionary learning loop iteratively optimizes alignment precision and motion smoothness, with the RL agent’s reward function designed to prioritize twist-angle accuracy and lateral displacement control. Additionally, a real-time monitoring sub-algorithm based on wavefront phase analysis tracks Newton’s ring evolution during stacking, providing feedback to the motion control module to mitigate layer misalignment.

\section*{\label{sec:level1}Data Availability}

The data that support the findings of this study are available upon reasonable request to the corresponding authors. We name our system as ``$\alpha$-Twist'', with a website available at www.alpha-twist.ai.

\section*{\label{sec:level2}Code Availability}

The code that support the findings of this study are available upon reasonable request to the corresponding authors.

\section*{\label{sec:level3}Acknowledgements}
This work is supported by the National Key R$\&$D Program of China (No. 2022YFA1203903) and the National Natural Science Foundation of China (NSFC) (Grant Nos. 12450003, 92565302, 92265203, 22325303, 62375160, 62274180). Z.V.H. acknowledges the support of the Fund for Shanxi “1331 Project” Key Subjects Construction, and the Innovation Program for Quantum Science and Technology (grant no. 2021ZD0302003). K.W. and T.T. acknowledge support from the JSPS KAKENHI (Grant Numbers 21H05233 and 23H02052) , the CREST (JPMJCR24A5), JST and World Premier International Research Center Initiative (WPI), MEXT, Japan. N.W. acknowledges the open project of key Laboratory of Artificial Structures and Quantum Control (Ministry of Education), Shanghai Jiao Tong University, and the Postgraduate Research $\&$ Practice Innovation program of Jiangsu province KYCX22$\_$0228. X.S. acknowledges support from Liaoning Provincial Natural Science Fund with Grant 2025-MS-053.

\section*{Author Contributions}
Z.H., W.H., J.Z.(jzhang74@sxu.edu.cn), J.L.(jmlu@lam.ln.cn), and X.L. conceived the experiment and supervised the overall project. X.L., J.H. and S.L. performed the device fabrications and low-frequency electrical measurements; K.Z., X.F., X.W., X.S., Z.L., C.Y., L.H., J.X., W.P., K.Y., and J.Z.(zj1752328218@gmail.com) performed the image marking for the machine-vision model trainings. Z.X., and T.Z. participated in hardware assembling. J.L.(jmlu@lam.ln.cn) contributed to device fabrications; K.W., T.T., M.T., and N.W. provided high quality h-BN bulk crystals; W.H., H.L., X.L., Z.W., and J.L.(jingli0107@xmu.edu.cn) performed the AI-algorithms. Z.H., X.L., J.H., and J.L.(jmlu@lam.ln.cn) analysed the experimental data. The manuscript was written by Z.H., X.L., and J.H. with discussion and inputs from all authors.

\section*{Competing Interests}
The authors declare no competing interests.

\setcounter{figure}{0}
\renewcommand{\figurename}{Extended Data Figure}

\newpage
\begin{figure*}[t!]
	\centering
	\includegraphics[width=0.9\linewidth]{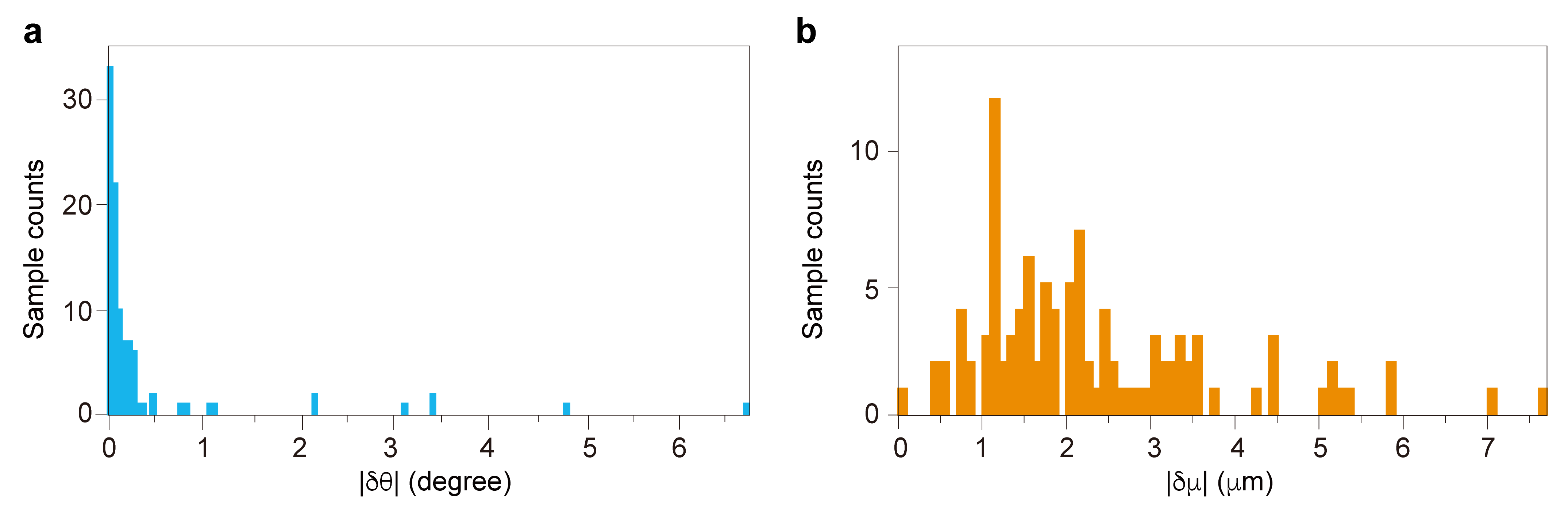}
	\caption{\textbf{Statistical evaluation of fabrication precision.}
		(a) Distribution of the absolute twist-angle error over a batch of $\sim 100$ assemblies, dominated by events below $0.3^\circ$ with a small tail above $0.3^\circ$.
		(b) Corresponding statistics of the lateral dislocation error for the same assembly batch.}
\end{figure*}
\clearpage

\newpage
\begin{figure*}[t!]
	\centering
	\includegraphics[width=0.9\linewidth]{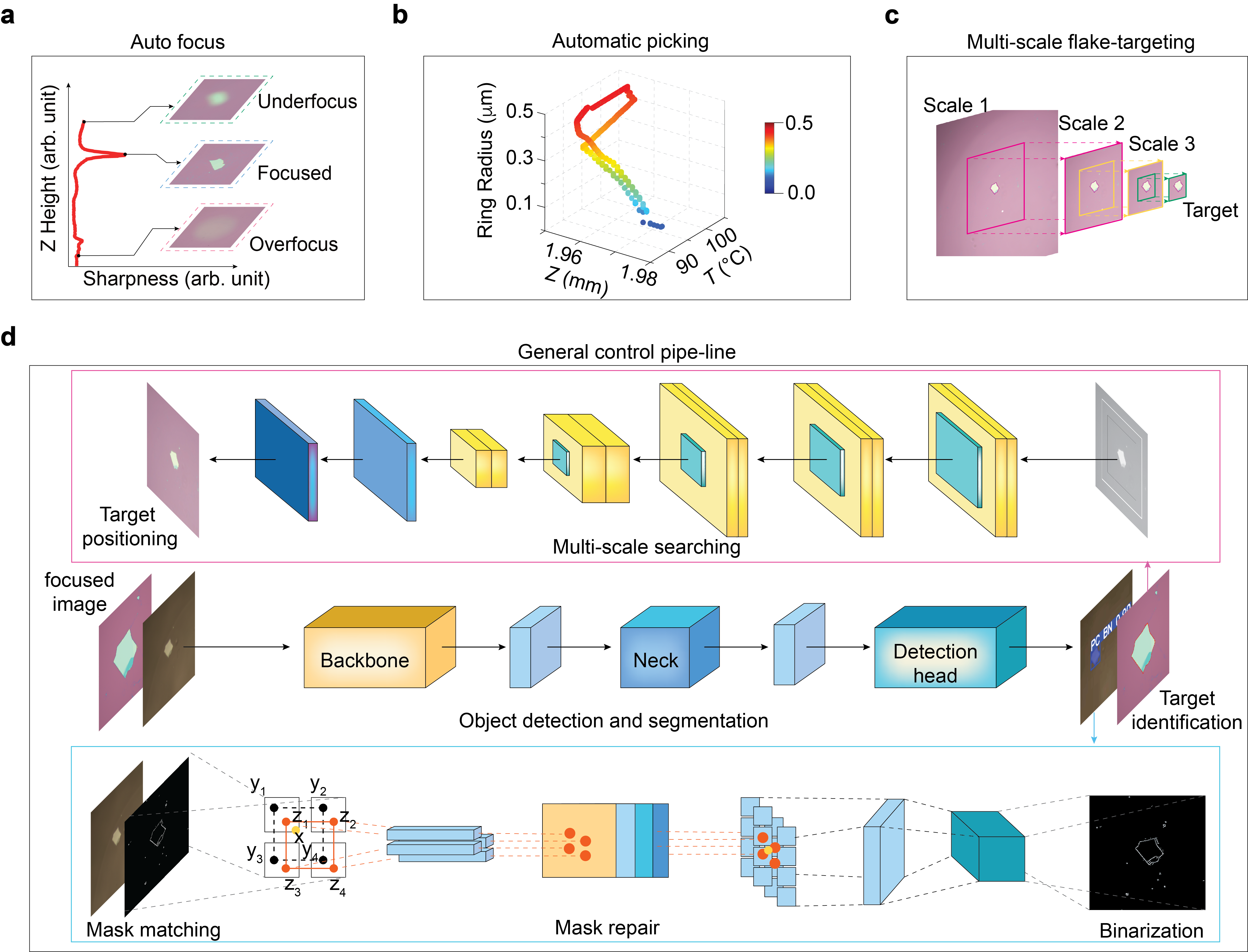}
	\caption{\textbf{Pipeline of the AI algorithm.} (a) Auto focus. (b) Automatic picking logged in parameter space of Newton's ring radius, stamp height, and temperature. (c) Multi-scale search strategy. (d) Diagram of the algorithms for target identification as well as multi-scale searching.}
\end{figure*}

\newpage
\begin{figure*}[t!]
	\centering
	\includegraphics[width=0.9\linewidth]{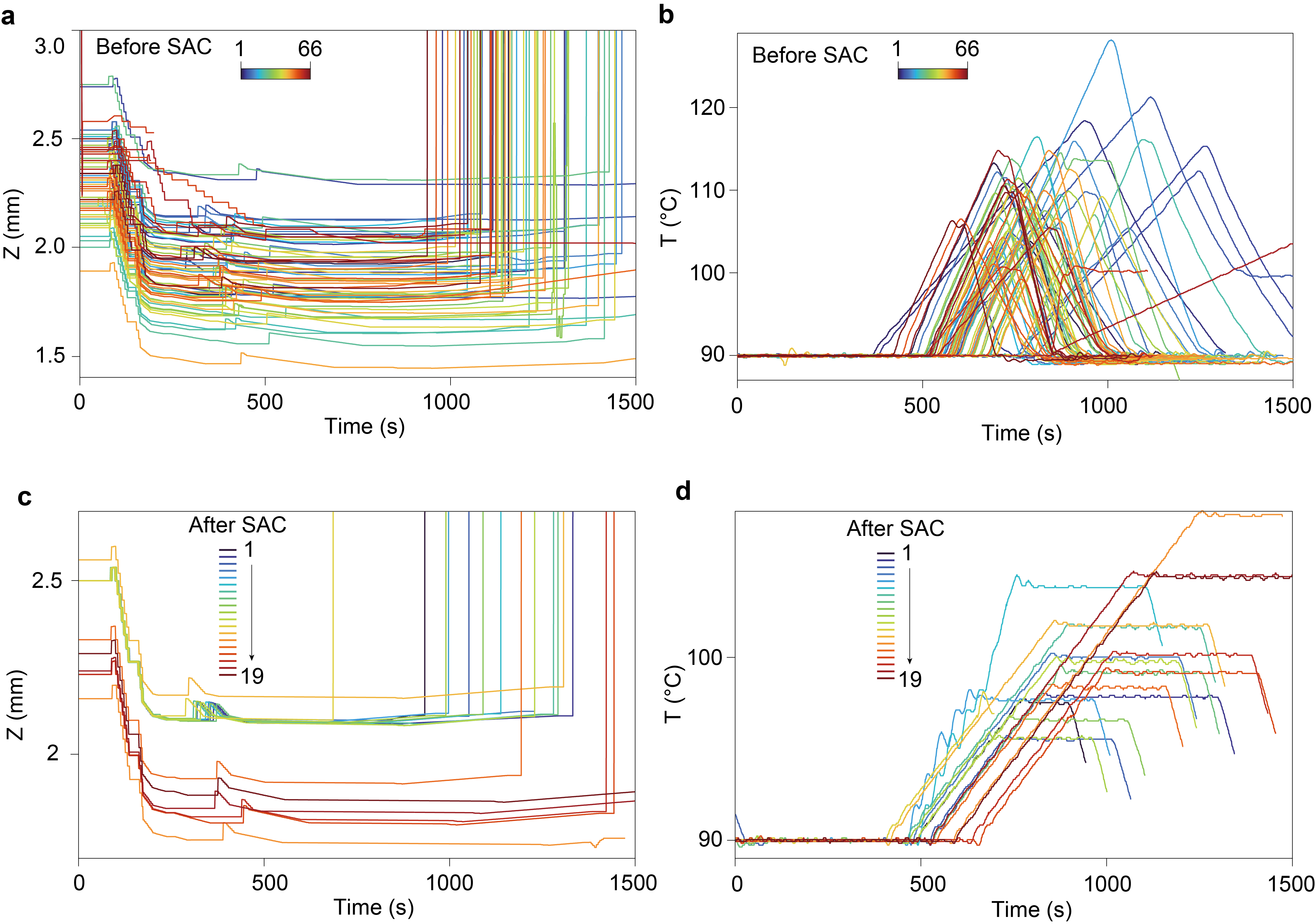}
	\caption{\textbf{The logged data of $Z$ and $T$ before and after SAC method.} (a) and (b) before SAC. (c) and (d) after SAC.}
\end{figure*}
\clearpage

\newpage
\begin{figure*}[t!]
	\centering
	\includegraphics[width=0.6\linewidth]{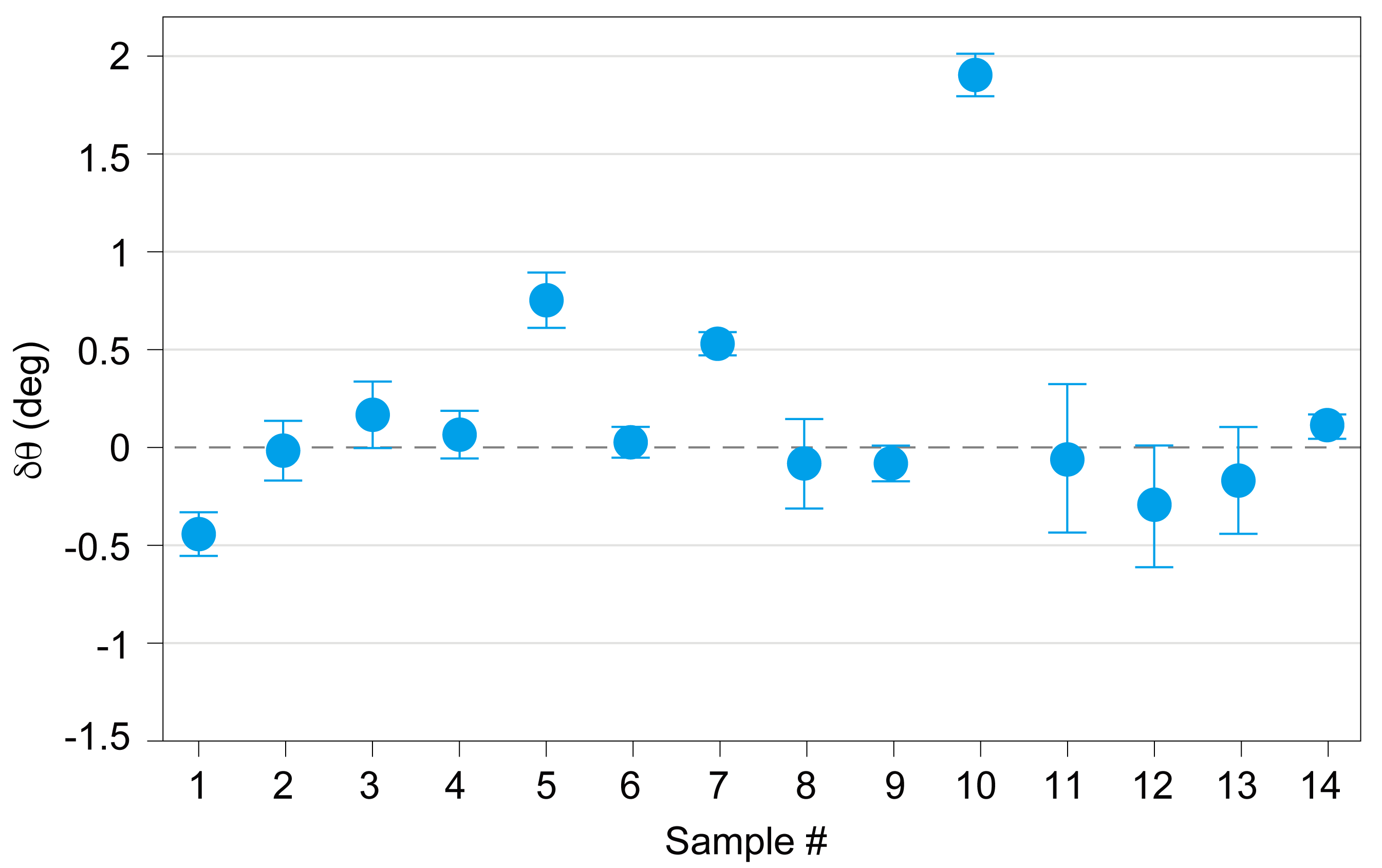}
	\caption{\textbf{Statistics of 14 TBLG stacks fabricated with a target angle of 1.1$^{\circ}$.} The deviation of twist angles from the AFM results is generally within $\pm$0.5$^{\circ}$.}
\end{figure*}
\clearpage

\newpage
\begin{figure*}[t!]
	\centering
	\includegraphics[width=0.7\linewidth]{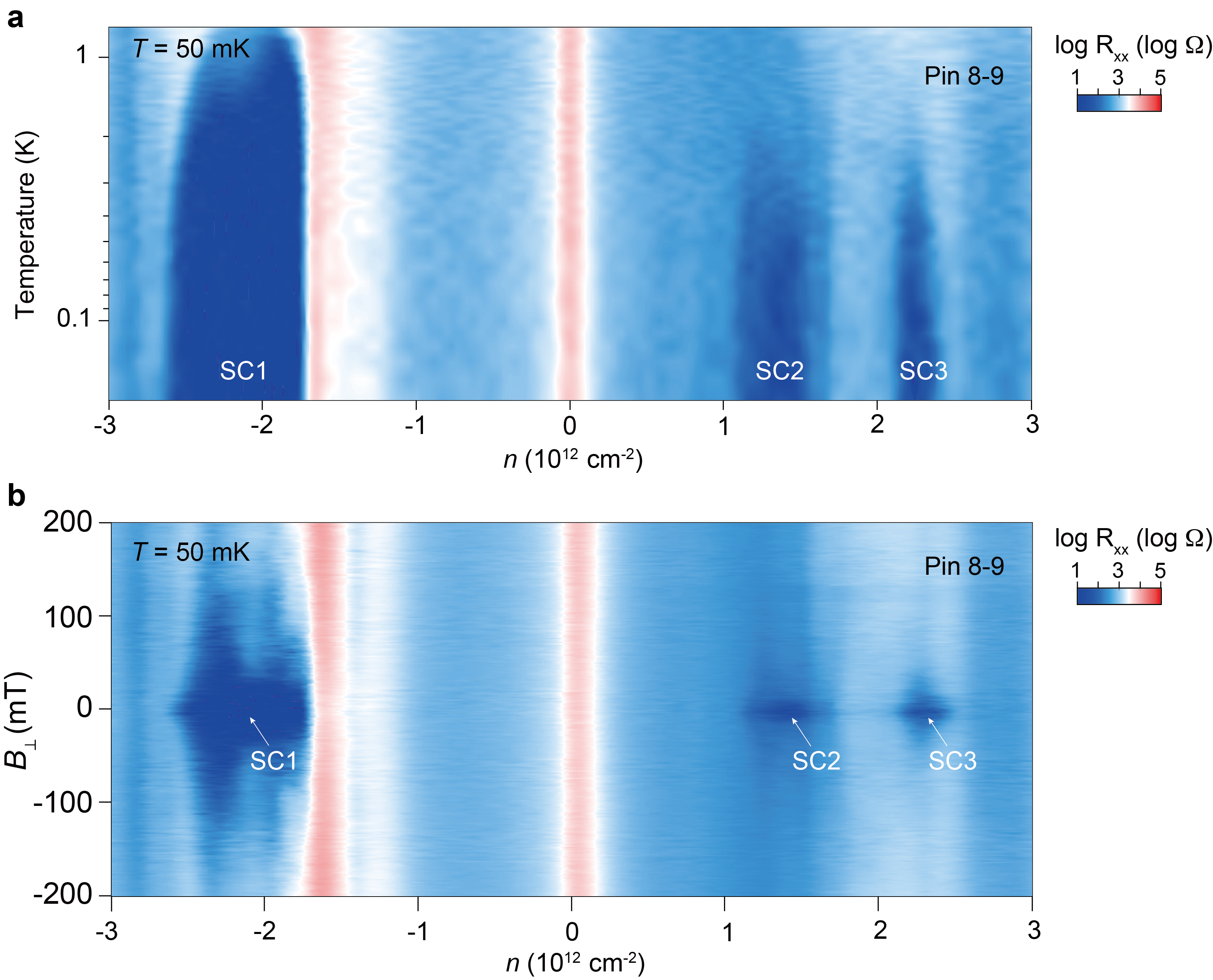}
	\caption{\textbf{Superconductivity in the robot-made TBLG device.} (a) Color map of longitudinal channel resistance $R_{\textrm{xx}}$ versus carrier density $n$ at different temperatures. (b) $R_{\textrm{xx}}$ versus $n$ at different perpendicular magnetic fields. Three superconducting pockets can be seen, with SC1 located at the hole side, and SC2-3 located at the electron side. Data measured in pins 8-9 of Device S-1.3-1.}
\end{figure*}

\newpage
\begin{figure*}[t!]
	\centering
	\includegraphics[width=0.9\linewidth]{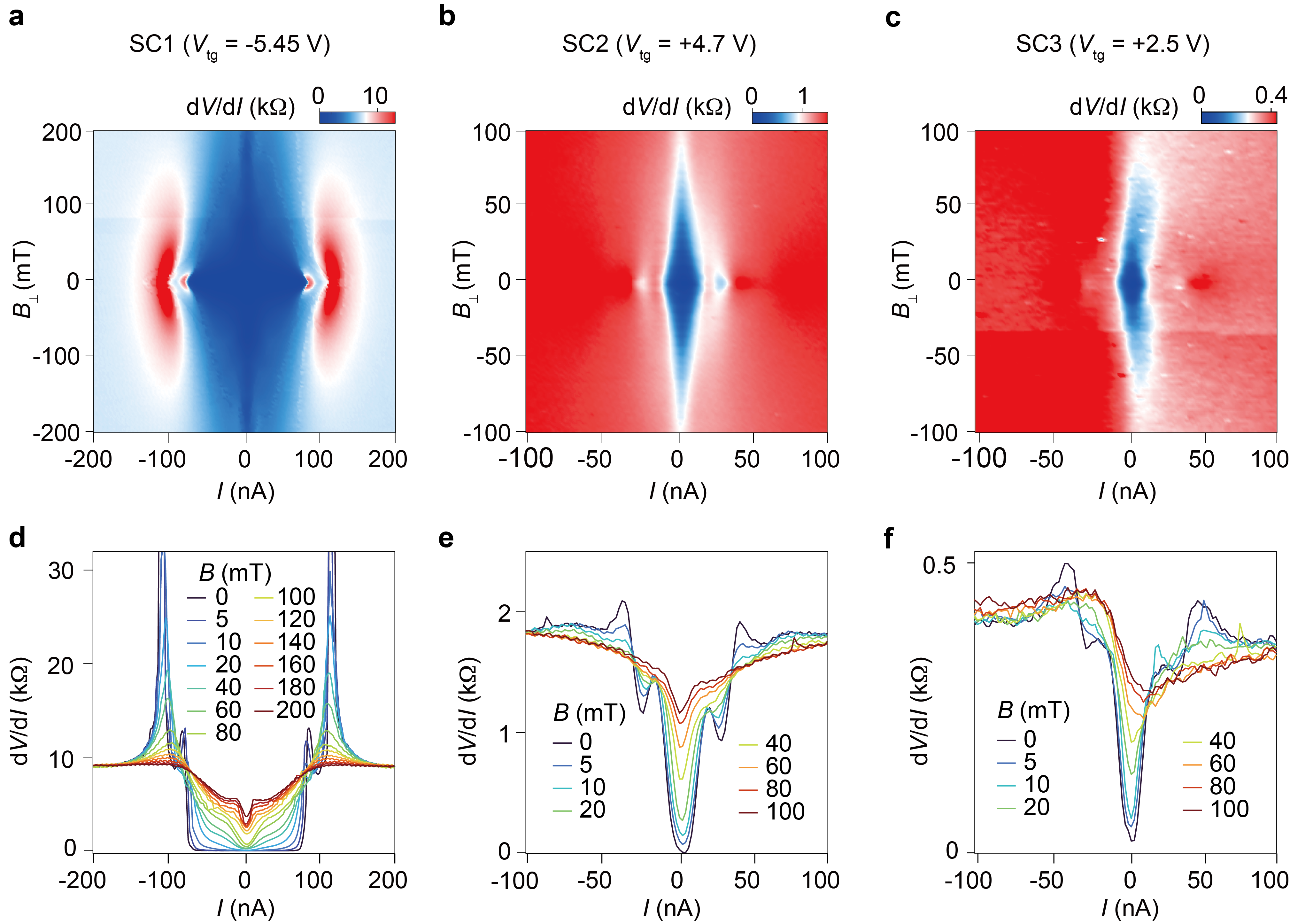}
	\caption{\textbf{Differential resistance in the superconducting pockets.} Measured across the same contact pins of the device shown in Extended Data Figure 3, color mappings of d$V$/d$I$ as a function of $I_{\textrm{bias}}$ and $B_{\perp}$ at $V_{\textrm{tg}}$ = -5.45 V, +4.7 V and + 2.5 V are illustrated in (a), (b) and (c), respectively. Line profiles of d$V$/d$I$ - $I_{\textrm{bias}}$ at different $B_{\perp}$ for each color map are plotted in (d), (e) and (f), respectively.}
\end{figure*}

\newpage
\begin{figure*}[t!]
	\centering
	\includegraphics[width=0.7\linewidth]{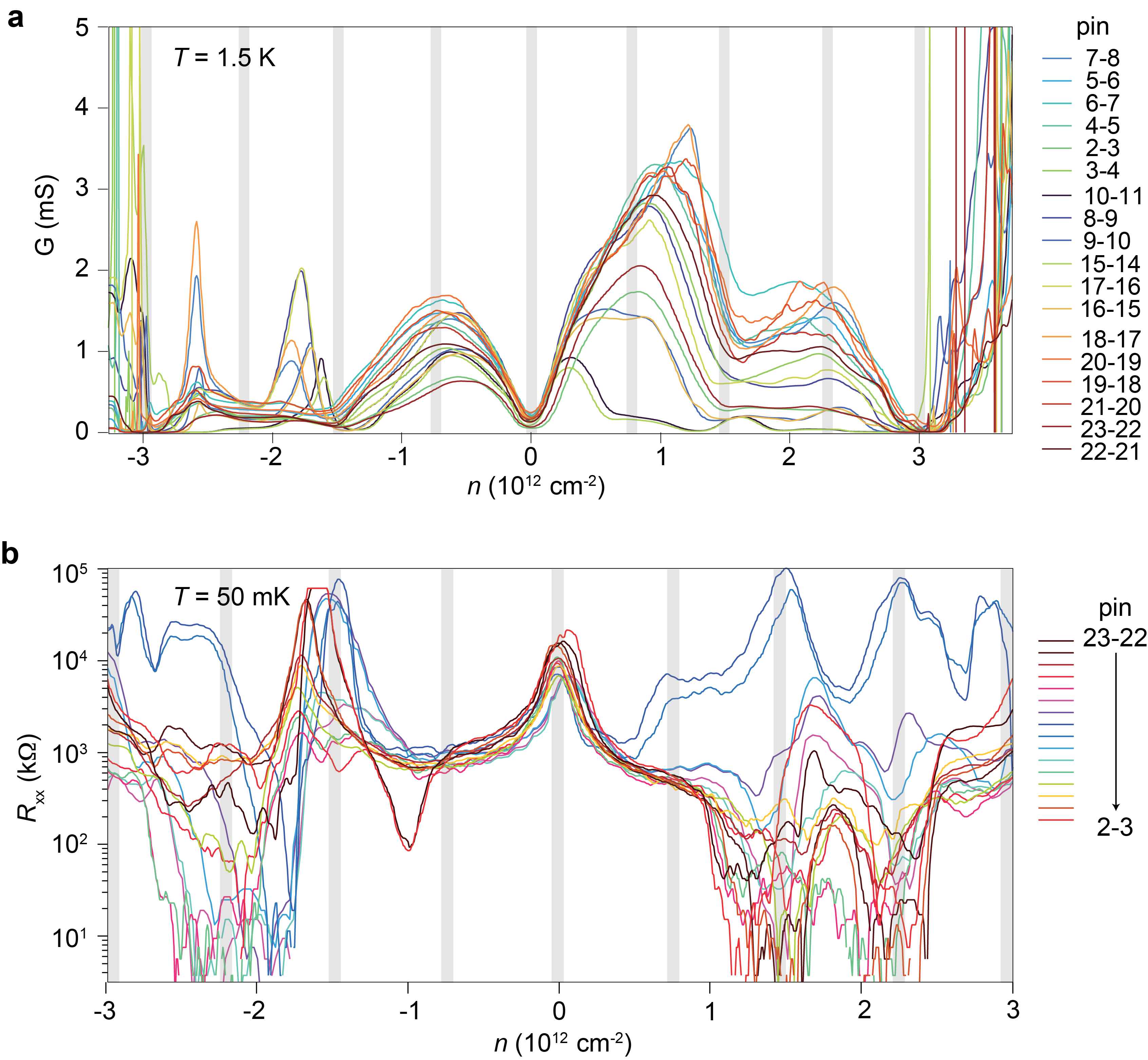}
	\caption{\textbf{Field effect curves measured across different pairs of contacts.} (a) Channel conductance as a function of carrier density for different pairs measured at $T$ = 1.5 K. (b) Channel resistance as a function of carrier density for different pairs measured at $T$ = 50 mK.}
\end{figure*}

\end{document}